\documentclass[journal=jctcce,manuscript=article]{achemso} 

\usepackage[version=3]{mhchem} 

\usepackage{graphicx}
\usepackage{dcolumn}
\usepackage{bm}
\usepackage[utf8]{inputenc}
\usepackage[T1]{fontenc}
\usepackage{mathptmx}
\usepackage{amsmath,amssymb}
\usepackage{bm}
\usepackage{cleveref}
\usepackage{booktabs}
\usepackage{notes2bib}
\usepackage{braket}
\usepackage{xcolor}
\usepackage[english]{babel}
\usepackage{siunitx}
\usepackage{notes2bib}

\usepackage{verbatim}

\makeatletter
\def\@email#1#2{%
 \endgroup
 \patchcmd{\titleblock@produce}
 {\frontmatter@RRAPformat}
 {\frontmatter@RRAPformat{\produce@RRAP{*#1\href{mailto:#2}{#2}}}\frontmatter@RRAPformat}
 {}{}
}%

\DeclareUnicodeCharacter{2212}{-}
\DeclareUnicodeCharacter{2061}{}

\def\br{\mathbf{r}}

\def\h2o{\mathrm{H}_2\mathrm{O}}

\def\icomp{\mathfrak{i}}

\author{Arno Förster}
\email{a.t.l.foerster@vu.nl}
\affiliation{Theoretical Chemistry, Vrije Universiteit Amsterdam, De Boelelaan 1105, 1081 HV Amsterdam, The Netherlands}%

\title{Beyond quasi-particle self-consistent $GW$ for molecules with vertex corrections}

\keywords{$GW$, vertex}

\begin{document}

\begin{abstract}
We introduce the $\Sigma^{\text{BSE}}@L^{\text{BSE}}$ self-energy in the quasi-particle self-consistent $GW$ (qs$GW$) framework (qs$\Sigma^{\text{BSE}}@L^{\text{BSE}}$). Here, $L$ is the two-particle response function which we calculate by solving the Bethe-Salpeter equation with the static, first-order $GW$ kernel. The same kernel is added to $\Sigma$ directly. For a set of medium organic molecules, we show that including the vertex both in $L$ and $\Sigma$ is crucial. This approach retains the good performance of qs$GW$ for predicting first ionization potentials and fundamental gaps, while it greatly improves the description of electron affinities. Its good performance places qs$\Sigma^{\text{BSE}}@L^{\text{BSE}}$ among the best-performing electron propagator methods for charged excitations. Adding the vertex in $L$ only, as commonly done in the solid state community, leads to devastating results for electron affinities and fundamental gaps. We also test the performance of BSE@qs$GW$ and qs$\Sigma^{\text{BSE}}@L^{\text{BSE}}$ for neutral charge-transfer excitation and find both methods to perform similar. We conclude that $\Sigma^{\text{BSE}}@L^{\text{BSE}}$ is a promising approximation to the electronic self-energy beyond $GW$. We hope that future research on dynamical vertex effects, second-order vertex corrections, and full self-consistency will improve the accuracy of this method, both for charged and neutral excitation energies. 
\end{abstract}

\section{\label{sec:introduction}Introduction}
The $GW$ approximation (GWA) to Hedin's equations for the electronic self-energy\cite{ Hedin1965, Strinati1988, Aryasetiawan1998, Fetter2003, martin2016} is by now widely used to calculate molecular charged excitation energies.\cite{Reining2018, Blase2018, Golze2019, Marie2024b} Here, $G$ stands for the single-particle Green's function, and $W$ is the dynamically screened Coulomb interaction typically calculated within the random phase approximation (RPA).\cite{Macke1950, Gell-Mann1957, Nozieres1958} The GWA approximates the exchange-correlation self-energy of a many-body system by adding correlation in the form of dynamical screening to the bare Fock exchange.\cite{Reining2018} 

Since $G$ and $W$ depend on the knowledge of $G$ itself, the GWA defines a self-consistent set of equations. 
During the self-consistent optimization of the interacting $G$, spectral weight, measured by the quasi-particle (QP) renormalization factor $Z$, is transferred from the QP peak to satellite features.\cite{Faleev2004, bruneval_springer2014} Consequently, the GWA underestimates the polarization due to particle-hole excitations and therefore severely underestimates the screening of the electron interaction.\cite{bruneval_springer2014} In practice, this leads to overestimated QP energies and gaps, and a poor description of satellites.\cite{Holm1998, Kutepov2012, Kutepov2016, Grumet2018, Rohlfing2023} 

Replacing the interacting $G$ with an effective $G^{(0)}$, typically obtained from a Kohn--Sham (KS)\cite{Kohn1965} calculation, and treating the $GW$ self-energy as a perturbation offers a simple solution to this problem.\cite{Hybertsen1985} Since $Z=1$ by definition, the under-screening problem is avoided. Careful selection of the KS starting point results in more accurate molecular QP energies at lower cost.\cite{Knight2016, Caruso2016, Bruneval2021a, McKeon2022, Bruneval2024, Golze2020, Li2022a} However, this so-called $G_0W_0$ approach introduces an undesirable dependence on the KS reference.\cite{Blase2011, Faber2011, Bruneval2013, Knight2016, Bruneval2019a, Bruneval2024} An iterative update of the $G_0W_0$ eigenvalues largely removes the starting-point dependence for charged excitations,\cite{Blase2011} but not for optical excitations from the $GW$-Bethe-Salpeter equation (BSE) method.\cite{Strinati1988, Onida2002, Forster2022c, Kshirsagar2023} 

Quasi-particle self-consistent $GW$ (qs$GW$)\cite{Faleev2004,VanSchilfgaarde2006, Kotani2007} addresses this starting-point dependence by replacing the Dyson equation for $G$ by an effective single-particle problem, obtained through Hermitization of the self-energy. By self-consistent optimization of the effective single-particle Hamiltonian, one finds an optimal $G^{(0)}$ within the GWA.\cite{bruneval_springer2014} The self-consistency condition rigorously removes the dependence on the DFT starting point\bibnote{Similar to other electronic structure methods, for instance, Hartree--Fock (HF) theory, there can of course be multiple solutions corresponding to different local minima which may be reached from different starting points} for charged and optical excitations.\cite{Forster2021a, Gui2018a, Forster2022} qs$GW$ has been applied first to solids\cite{Faleev2004,VanSchilfgaarde2006, Kotani2007, Shishkin2007, Bruneval2006a} where it performs somewhat better than sc$GW$ but still overestimates the band gaps by about 10-20\%.\cite{VanSchilfgaarde2006, Shishkin2007, Chen2015, Tal2021, Cunningham2023} 

This issue can be addressed by including higher-order diagrams in the self-energy through so-called vertex corrections which describe the scattering between particle-hole pairs. The vertex enters the self-energy $\Sigma$ directly, but also through the response function $L$ which in Hedin's equations determines $W$.\cite{martin2016} Within qs$GW$, vertex corrections have been predominantly considered in $L$ only, but not in the self-energy.\cite{Shishkin2007, Chen2015, Cunningham2018, Cunningham2023} The reasons for this are two-fold. First, vertex-corrections in $L$, be it through approximate exchange-correlation kernels\cite{Shishkin2007, Chen2015} or the non-local first-order vertex in Hedin's equations,\cite{Cunningham2018, Cunningham2023} account for the electron-hole interaction missing in the RPA and therefore close the band gap, while corrections in the self-energy are known to open the gap.\cite{Gruneis2014, Bruneval2024} The second argument relies on the Ward identity which states that the vertex function $\Gamma$ goes as $1/Z$ in the long-range and zero-frequency limit.\cite{Kuwahara2014} Since in the exact self-energy $\Sigma = i GW\Gamma$ the $Z$-factors in $G$, which renormalizes $G^{(0)}$, and $\Gamma$ should cancel,\cite{Kutepov2016, Kutepov2017} Kotani, van Schilfgaarde and Faleev argued that vertex corrections in $\Sigma$ should be avoided whenever $G$ is replaced by $G^{(0)}$.\cite{Kotani2007} Pasquarello and coworkers argued that this argument only applies in the long-range and in addition to the full vertex in $L$, introduced a short-range vertex to the self-energy\cite{Tal2021} which they found to have a small effect on band gaps but to improve QP energies.\cite{Tal2021, Abdallah2024} 

qs$GW$ has also been used for atoms\cite{Bruneval2012} and molecules.\cite{Ke2011, Kaplan2016, Forster2021a} It gives relatively accurate ionization potentials (IP)\cite{Caruso2016, Marie2023a, Marie2024} and its fundamental gaps are in excellent agreement with $\Delta$-Coupled cluster (CC) with single, double and perturbative triple substitutions [$\Delta$CCSD(T)]\cite{Purvis1982} for the ACC24 of 24 organic acceptor molecules\cite{Knight2016} relevant for photovoltaic applications.\cite{Forster2022} Interestingly, qs$GW$ tends to underestimate these systems' fundamental gaps\cite{Forster2022} and the inclusion of a vertex correction in $L$ but not in $\Sigma$ can only be expected to worsen them. Vertex corrections in $L$ only have been tried in molecular $G_0W_0$ calculations with HF starting points. While they often improve first IPs,\cite{Lewis2019, Bruneval2021a, Forster2024} they also produce major outliers in benchmarks and are therefore unreliable.

Only a few vertex-corrected qs$GW$ calculations have been performed for molecules, and only vertex corrections in the self-energy have been tried.\cite{Forster2022, Forster2023a, Bruneval2024} These calculations either follow \citet{Gruneis2014} and perturbatively correct the qs$GW$ QP energies using a statically screened version\cite{Gruneis2014, Forster2022} of the $G3W2$ self-energy,\cite{Kutepov2016, Bruneval2024} or use the fully dynamical $G3W2$ correction to $GW$.\cite{Bruneval2024} In all cases, no clear improvements over qs$GW$ could be observed. This is not surprising since perturbative vertex corrections to the self-energy\cite{Ren2015, Wang2021, Forster2022, Bruneval2024} are a bad strategy in molecular $G_0W_0$ calculations, and can only work if a qualitatively incorrect $G^{(0)}$ is used as a starting point:\cite{Bruneval2024} for a PBE or PBE0 $G^{(0)}$, the GWA systematically underestimates the first IPs.\cite{VanSetten2015, Caruso2016, Knight2016} Vertex corrections in $\Sigma$ only will increase the first IPs and therefore improve the final results.\cite{Ren2015, Wang2021, Wen2024} Since qs$GW$ is already relatively accurate, no clear improvements can be expected. Only the statically screened $G3W2$ correction retains the good performance of qs$GW$, but only because its magnitude is small.\cite{Forster2022}

Starting from a Hartree-Fock (HF) Green's function, solving the time-dependent HF equations for $L$, and using the same vertex in $\Sigma$ in $G_0W_0$ has been a much more successful strategy.\cite{Maggio2017, Maggio2018, Vlcek2019, Mejuto-Zaera2021, Vacondio2024, Patterson2024, Patterson2024a, Forster2024} This scheme performs well for the calculation of first IPs of atoms,\cite{Vacondio2024},smaller molecules in the GW100 set\cite{Forster2024} and also larger systems like linear acenes\cite{Vlcek2019} and others.\cite{Patterson2024} 
Patterson [using the Tam-Dancoff approximation (TDA)]\cite{Patterson2024, Patterson2024a} and \citet*{Forster2024} additionally replaced the bare exchange in the HF equations with statically screened exchange. This replacement is crucial for realistic charged excitations in larger systems where screening effects are important.\cite{Forster2024}

Here, we use the same self-energy in molecular qs$GW$ calculations and report benchmark results for IPs, EAs, and fundamental gaps for the ACC24 set. While the more diverse GW100 set\cite{VanSetten2015} is often used to benchmark different $GW$ approximations, we use here the ACC24 set since it also allows us to assess the accuracy for electron affinities (EA) and fundamental gaps. Furthermore, the molecules in ACC24 are representative of the medium organic systems to which (vertex-corrected) $GW$ is often applied.\cite{Wang2020, Stuke2020, Bhattacharya2024} Since we solve the BSE with the screened exchange vertex self-consistently and not only after the $GW$ calculation, we also investigate how the neutral (singlet) excitation energies compare to the ones obtained from standard BSE@qs$GW$ for the QUEST \#6 subset\cite{Loos2021} of charge-transfer excitations (CT) of the QUEST database.\cite{Loos2018b, Loos2020f, Veril2021} In section~\ref{sec:Theory}, we introduce our expression for the self-energy and discuss differences to the closely related method of \citet{Cunningham2023} In section~\ref{sec:Results} we present and discuss our numerical results and conclude this work in section~\ref{sec:Conclusions}.

\section{\label{sec:Theory}Theory}
The exact self-energy of an interacting many-electron system can be obtained by solving the following set of coupled equations self-consistently:
\begin{align}
\nonumber
\Sigma_{xc}(1,2) = & \icomp v(1^+,2) G(1,2) \\
\label{sigma}
  & + \icomp v(1^+,3) G(1,4) I(4,6,2,5) L(5,3,6,3) \\
  \nonumber
L(1,2,3,4) = & L^{(0)}(1,2,3,4) \\
  \label{response}
    & + L^{(0)}(1,5,3,6) I (6,7,5,8)
    L(8,2,7,4) \\
    \label{dyson}
    G(1,2) = & G^{(0)}(1,2) + G(1,4)\Sigma_{xc}(3,4) G(4,2) \;.
\end{align}
\Cref{sigma} combines the 2-point Coulomb interaction $v$, the 2-particle correlation function $L$, the single-particle Green's function $G$, and the kernel 
\begin{equation}
\label{kernel}
    I(1,2,3,4) = \delta(1,3)\delta(2,4)v(1,4) + \icomp \frac{\delta \Sigma_{xc}(1,3)}{\delta G (4,2)} \;.
\end{equation}
The same kernel also appears in the Bethe-Salpeter equation (BSE) \cref{response} for $L$, where $L^{(0)}(1,2,3,4) = -\icomp G(1,4)G(2,3)$ is the non-interacting 2-particle correlation function. Integration over repeated indices is implied and integers $n = (\bm{r}_n, \sigma_n, t_n)$ collect spatial coordinates, spin, and time. The same or slightly similar sets of equations are known at least since the work of \citet{Baym1961} and are frequently encountered in the literature.\cite{Romaniello2012, Maggio2017, martin2016, Orlando2023, Vacondio2024, Forster2024, Patterson2024} They are completely equivalent to Hedin's equations.\cite{Strinati1988, Starke2012, Maggio2017} 

\subsection{quasi-particle self-consistent $GW$}
In this work, we solve this set of equations in the qs$GW$ approximation. The aim of a qs$GW$ calculation\cite{Faleev2004, VanSchilfgaarde2006, Kotani2007, bruneval_springer2014} is to find an effective $G^{(0)}$ which approximates the interacting $G$ in \cref{dyson}.\cite{Kotani2007} Assuming the self-energy is constant in the vicinity of some reference QP energy $\epsilon_p$, we rewrite \cref{dyson} as
\begin{equation}
\label{dyson_general_linearized2}
     \sum_{q}\left[\Sigma_{pq}(\epsilon_p) + \epsilon_p\delta_{pq} \right] \phi_{q} = \omega \phi_{p} \;.
\end{equation} 
To get an effective $G^{(0)}$, the self-energy has to be mapped to an effective Hermitian QP Hamiltonian. Different mappings have been proposed\cite{Faleev2004, Kotani2007, Shishkin2007, Sakuma2009, Kutepov2017b, Kuwahara2014, Marie2023a} and we follow Ref.~\citenum{Kotani2007} and set
\begin{equation}
\label{QP-Hamiltonian}
    H_{pq} = \frac{1}{2} \left\{ 
    \text{Re} \Sigma_{pq}(\epsilon^{QP}_p) + 
    \text{Re} \Sigma_{pq}(\epsilon^{QP}_q)
    \right\} \;.
\end{equation}
This construction of the QP Hamiltonian has been shown to satisfy a variational principle.\cite{Ismail-Beigi2017} For reasons of numerical stability, a closely related construction\cite{Faleev2004} which evaluates the off-diagonal elements of $\hat{H}^{QP}$ at the Fermi-level instead is often preferred in implementations which calculate $\Sigma$ by analytical continuation.\cite{Forster2021a, Lei2022, Harsha2024} Comparisons between different QP Hamiltonians show that they typically give similar QP energies.\cite{Forster2021a, Marie2023a} This also includes linearized qs$GW$ where the self-energy is Taylor-expanded around the chemical potential and the linear term is retained.\cite{Kutepov2017b, Shishkin2007, Kuwahara2014, Harsha2024} In conclusion, we expect all of our findings to be also valid for other QP Hamiltonians.

Generally, qs$GW$ is expected to work well if QP renormalization is weak, i.e. if there is a single dominant QP peak. For the organic molecules contained in the ACC24 set for which we will present results below this is always the case.\cite{Dolgounitcheva2016} This assumption is however not generally valid and in cases where QP renormalization is strong, full self-consistency will be important.\cite{Harsha2024} qs$GW$ is inherently limited to QP peaks, and full self-consistency is also necessary to recover any other spectral features beyond $G_0W_0$. Furthermore, at least for extended systems, vertex corrections are known to behave qualitatively differently in sc$GW$ and qs$GW$.\cite{Kutepov2022} The extension of the self-energy approximations of Ref.~\citenum{Forster2024} to sc$GW$ is outside the scope of this work but will be eventually addressed in future work. Promising work in this direction has recently been published by Zgid and coworkers.\cite{Pokhilko2024a}

\subsection{Vertex-corrections in quasi-particle self-consistent $GW$}
\Cref{sigma,response,dyson_general_linearized2} form a closed set of equations. In each iteration, diagonalization of \cref{dyson_general_linearized2} yields a set of molecular orbitals $\varphi_p$ and QP energies $\epsilon_p$ from which the non-interacting $G^{(0)}$ is constructed:
\begin{equation}
  \label{eq:g0}
  G^{(0)}(\br,\br',\omega) =
    \sum_i^\mathrm{occ}
     \frac{\varphi_i(\br)   \varphi_i(\br^\prime)}
     {\omega - \epsilon_i -\icomp \eta }
+ \sum_a^\mathrm{virt}
     \frac{\varphi_a(\br)   \varphi_a(\br^\prime)}
     {\omega - \epsilon_a +\icomp \eta } \;.
\end{equation}
We now approximate the kernel \cref{kernel} as
\begin{equation}
\label{kernel_approximation}
    I(1,2,3,4) \approx \delta(1,3)\delta(2,4)v(1,4) - \delta(1,4)\delta(3,2)W_0(1,3) \;,
\end{equation}
where $W_0$ is the statically screened Coulomb interaction in the (direct) RPA. Formally, \cref{kernel_approximation} is obtained by making the GWA to the self-energy in \eqref{kernel}, neglecting the variation of the dynamical $W$ with respect to $G$ and taking the static limit.
With \cref{kernel_approximation,eq:g0}, \cref{response} becomes a function of a single frequency and can be solved exactly by diagonalization in the particle-hole representation. In the usual notation, \cref{response} becomes\cite{Sander2015}
\begin{equation}
\label{BSE}
    \begin{pmatrix}
    \bm{A} & \bm{B} \\ 
    \bm{B} & \bm{A} \\
    \end{pmatrix}
    \begin{pmatrix}
    \bm{X}\\ 
    \bm{Y}\\
    \end{pmatrix} 
    = 
\begin{pmatrix}
    \bm{\Omega} & \bm{0} \\ 
    \bm{0} & -\bm{\Omega} \\    
\end{pmatrix}
    \begin{pmatrix}
    \bm{X}\\ 
    \bm{Y}\\
    \end{pmatrix} 
\end{equation}
with the matrix elements 
\begin{equation}
\label{amt_elements}
\begin{aligned}
A_{ia,jb} = & - \delta_{ij} \delta_{ab}(\epsilon_a - \epsilon_i) + (ia|v|jb) - (ij|W_0|ab) \\
B_{ia,jb} = & (ia|v|bj) - (ib|W_0|ja)
\end{aligned}
\end{equation}
and chemists' notation for the 2-electron integrals,
\begin{equation}
\label{eq:eri}
( pq | K | rs ) = \int d\br \int d\br'
   \phi^*_p(\br) \phi_q(\br) K(\br,\br')
   \phi^*_r(\br') \phi_s(\br') \; .
\end{equation}
$\bm{\Omega}$ is a diagonal matrix containing the system's neutral excitation energies.
Solving only for the particle-hole part constitutes no additional approximation beyond the approximation to the kernel \cref{kernel_approximation}. While $L$ generally also includes particle-particle and hole-hole terms, these contributions are eliminated when the first-order kernel $I$ \cref{kernel_approximation} is static.\cite{Romaniello2009b} In the basis of molecular orbitals which diagonalize \cref{dyson_general_linearized2}, the correlation part of the self-energy \cref{sigma} with the very same kernel \cref{kernel_approximation} can then be shown to be\cite{Forster2024, Vacondio2024}
\begin{equation}
\label{sigmFull}
\Sigma_{pq}(\omega) = \Sigma^<_{pq}(\omega) + \Sigma^>_{pq}(\omega)
\end{equation}
with 
\begin{equation}
\label{sigmFullgreater}
\begin{aligned}
 \Sigma^<_{pq}(\omega)= & \sum_S \sum^{occ}_k \frac{1}{\omega - \epsilon_k + \Omega_S - i \eta} \\
  & \times
 \left[\sum_{ia}  2( a i | v | q k ) ( X_{ia}^S  + Y_{ia}^S ) -
 ( k a| W_0 | q i ) X_{ia}^S 
               - ( k i | W_0 | q a ) Y_{ia}^S
  \right] \\
  & \times
   \left[ \sum_{jb} ( b j | v | p k ) ( X_{jb}^S  + Y_{jb}^S )
  \right]
    \end{aligned}
\end{equation}
and
\begin{equation}
\label{sigmFulllesser}
\begin{aligned}
\Sigma^>_{pq}(\omega) = & \sum_S \sum^{virt}_c \frac{1}{\omega - \epsilon_c - \Omega_S
    + i \eta} \\
 & \times
 \left[\sum_{ia} 2( a i | v | q c ) ( X_{ia}^S  + Y_{ia}^S )
 -( c i | W_0 | q a ) X_{ia}^S 
                - ( c a | W_0 | q i ) Y_{ia}^S
  \right] \\
  & \times
   \left[ \sum_{jb} ( b j | v | p c  ) ( X_{jb}^S  + Y_{jb}^S ) 
  \right] \;.
  \end{aligned}
\end{equation}
We start our vertex-corrected qs$GW$ calculations from a Hartree-Fock reference. We then solve the RPA equations using the HF $G^{(0)}$ to obtain $W_0$ which we use to solve \cref{BSE}. Afterward, the self-energy \cref{sigmFull} is constructed and used in \cref{QP-Hamiltonian}. We then diagonalize \cref{dyson_general_linearized2} to obtain a new set of orbitals $\phi_k$ and QP energies $\epsilon_k$ and construct the corresponding $G^{(0)}$ \cref{eq:g0}. This process is iterated until convergence. Since the final results will not depend on the mean-field reference, one could also initialize the self-consistency cycle with a KS reference.

following our recent work,\cite{Forster2024} we refer to the self-energy in \cref{sigmFull,sigmFullgreater,sigmFulllesser} as $\Sigma^{BSE}@L^{BSE}$. For $W_0 = 0$, \cref{sigmFull} reduces to the GWA with BSE screening ($GW@L^{BSE}$), and for $W_0 = 0$ also in \cref{amt_elements}, to the GWA. The former variant has been coined qs$G\hat{W}$ by Cunningham \textit{et. al.}\cite{Cunningham2018, Cunningham2023} and been applied to a variety of materials.\cite{Acharya2021, Acharya2021a, Radha2021, Dadkhah2023, Garcia2024} They solved \cref{BSE} within the TDA\cite{Hirata1999} which has later been shown to be a rather severe approximation.\cite{Kutepov2022} 
The TDA has been shown to decrease the first IPs of furan and diacetylene at the $\Sigma^{TDHF}$ level of theory by 0.21 and 0.29 eV respectively.\cite{Waide2024}
The qs$G\hat{W}$ of Cunningham \textit{et. al.}\cite{Cunningham2018, Cunningham2023} should not be confused with the qs$G\hat{W}$ method of \citet*{Tal2021}.

Our work extends the method by \citet{Cunningham2023} in two important aspects. First, we do not make the TDA in the solution of the BSE [which would correspond to $\bm{B} = 0$ in \cref{BSE}]. Second, and more importantly, we include the very same kernel as in \cref{BSE} also in the self-energy. Due to its Dyson-like structure, the BSE with the matrix elements \cref{amt_elements} generates particle-hole ladder diagrams and resumms them to all orders. They describe the electron-hole attraction missing in the RPA and therefore close the band gap.\cite{Cunningham2023} This is not desired in molecular qs$GW$ calculations since band gaps are already underestimated.\cite{Forster2022} At the $G_0W_0$@HF level, it has already been pointed out before that going beyond the RPA for the screening often deteriorates the molecular IPs,\cite{Lewis2019} and should be combined with vertex corrections in $\Sigma$.\cite{Maggio2017, Patterson2024} As shown in Refs.~\citenum{Forster2024}vertex corrections in the self-energy come with an opposite sign and are needed to balance the vertex correction in the screening. Likewise, the importance of vertex corrections for accurate EAs has been pointed out in Ref.~\citenum{Vlcek2019}.

It is worthwhile to point out certain connections to the $T$-matrix approximations to the electronic self-energy which have recently been used to calculate molecular IPs.\cite{Zhang2017, Monino2023, Marie2024} 
One can see that the terms beyond $GW$ in our self-energy expression are precisely the ones in the electron-hole $T$-matrix self-energy when one only considers the exchange term therein (compare for instance to eqs. 43-44 in Ref.~\citenum{Orlando2023}). However, \cref{sigmFullgreater,sigmFulllesser} are not just the sum of the GWA and the electron-hole $T$-matrix (minus a double counting correction which would be obtained from counting the direct term twice which is both present in the GWA and the $T$-matrix). In the electron-hole $T$-matrix, the 2-particle correlation function $L$ is obtained from the "exchange" version of the usual RPA.\cite{Orlando2023} Here, $L$ is calculated from the BSE which contains both direct and exchange interactions, combining the usual direct RPA and "exchange"-RPA. The connection becomes even more apparent if one considers TDHF instead of the BSE, obtained by replacing $W_0$ with $v$ in \cref{amt_elements}. 

Our work does not include the particle-particle $T$-matrix since no particle-particle ladders are present in $L$. The particle-particle $T$-matrix diagrams can be obtained Within Hedin's scheme from higher-order vertex corrections obtained as functional derivatives of $W$ and the three-point vertex $\Gamma$.\cite{Romaniello2012, Mejuto-Zaera2022a, Cunningham2024}It is also possible to start from the particle-particle $T$-matrix and add $GW$ diagrams as vertex corrections.\cite{Marie2024a} Combining the GWA with both $T$-matrix approximations would correspond to the fluctuation exchange (FLEX) approximation.\cite{Bickers1989, Gukelberger2015}

\subsection{Computational Details}

We performed vertex-corrected qs$GW$ calculations with a development version of the BAND engine\cite{TeVelde1991} of the Amsterdam modeling suite (AMS2024), using the analytical frequency integration expression for the self-energy following Refs.~\citenum{Bruneval2016a} and~\citenum{Marie2023a}. 
We have performed all calculations with augmented Dunning basis sets of DZ to QZ quality (aug-cc-pVDZ to aug-cc-pVQZ).\cite{Dunning1989, Kendall1992} We discuss the basis set requirements of all calculations below.

All 4-center integrals are calculated using the pair-atomic density fitting scheme as presented in Ref.~\citenum{Spadetto2023}. The size of the auxiliary basis in this approach can be tuned by a single threshold which we set to $\varepsilon_{aux} = 1 \times 10^{-10}$ in all calculations if not stated otherwise. We further artificially enlarge the auxiliary basis by setting the BoostL option.\cite{Spadetto2023} We further eliminate almost linear dependent products of basis functions from the primary basis by setting the $K$-matrix regularization parameter to $5 \times 10^{-4}$.\cite{Spadetto2023}

Since BAND does not support molecular point group symmetry, we calculated neutral excitations with the ADF engine.\cite{Snijders2001,Forster2022c} For the QUEST \#6 database of molecular CT excitations\cite{Loos2021} we used STO basis sets ranging from TZP and TZ2P\cite{vanLenthe2003} to TZ3P.\cite{Forster2021} The latter is comparable to cc-pVTZ. Since we use reference values obtained with the cc-pVTZ basis set,\cite{Loos2021} the remaining basis set error should be negligible as well. We verified this by performing additional cc-pVTZ calculations with the BAND code for cases where the correct symmetry of the excitation can easily be verified (see SI for details). 

To converge the qs$GW$ and qs$\Sigma^{BSE}$ calculations for the ACC24 set, we use a linear mixing strategy,\cite{Forster2023a} in which we construct the QP Hamiltonian in \eqref{QP-Hamiltonian} for the $n$th iteration as $H^{(n)} \leftarrow \alpha^{(n)} H^{(n)} + (1-\alpha^{(n)})H^{(n-1)}$. 
We start the self-consistency field (SCF) cycle with $\alpha^{(0)} = 0.3$. In case the SCF error decreases, we use the mixing parameter $\alpha^{(n)} = \max \left\{1.2 \times \alpha^{(n-1)},0.5\right\}$ in the $n$th iteration. In case the SCF error increases, we reset the mixing parameter to $\alpha^{(0)}$. We terminate the calculation when the change in the fundamental gap between 2 subsequent iterations is smaller than 1 meV. For reasons of computational efficiency, for the molecules in the QUEST \#6 database we use the direct inversion of the iterative subspace (DIIS)\cite{Pulay1980} algorithm of Ref.~\citenum{Forster2021a} (which is based on Ref.~\citenum{Veril2018}) with a maximum of 8 iterations.

\section{\label{sec:Results}Results}

\subsection{Charged excitations}

\subsubsection{Basis set dependence}
The slow convergence of individual $GW$ QP energies to the complete basis set (CBS) limit in Gaussian basis sets is well-documented.\cite{Ke2011, Bruneval2012, Bruneval2013, Bruneval2020} It is however rarely emphasized that $GW$ QP energies converge slower to the basis set limit than the ones calculated with CC methods which are often used for benchmarking $GW$, like $\Delta$CCSD(T) or equation-of-motion (EOM)-CC\cite{Monkhorst1977, Koch1990, Stanton1993} methods. These different convergence rates imply that comparison to CC reference values should ideally be performed in a large basis set to eliminate basis set errors. Alternatively, the basis set limit may be estimated by extrapolation which requires performing at least a QZ calculation for a relatively precise estimate.\cite{Stuke2020} As a rule of thumb, the extrapolated result will be roughly of 5Z quality.\cite{Jensen2013} Since GTO-type basis sets do not converge uniformly to the complete basis set limit such techniques are not ideally suited for precise benchmarks. 

In the case of the ACC24 benchmark, \citet{Richard2016} report basis set limit extrapolated first IPs and EAs. The extrapolated values are always obtained by performing a $\Delta$CCSD(T) calculation using a series of calculations with aug-cc basis sets ranging from DZ to QZ quality. The remaining basis set error is estimated by extrapolation with the help of $\Delta$MP2 calculations in larger bases. However, for only 6 out of the 24 systems in their set, they performed a $\Delta$CCSD(T)/aug-cc-pvQZ calculation and for 4 of them, they only reported $\Delta$CCSD(T)/aug-cc-pvDZ calculations. 

\begin{figure}[hbt!]
    \centering
    \includegraphics[width=0.7\linewidth]{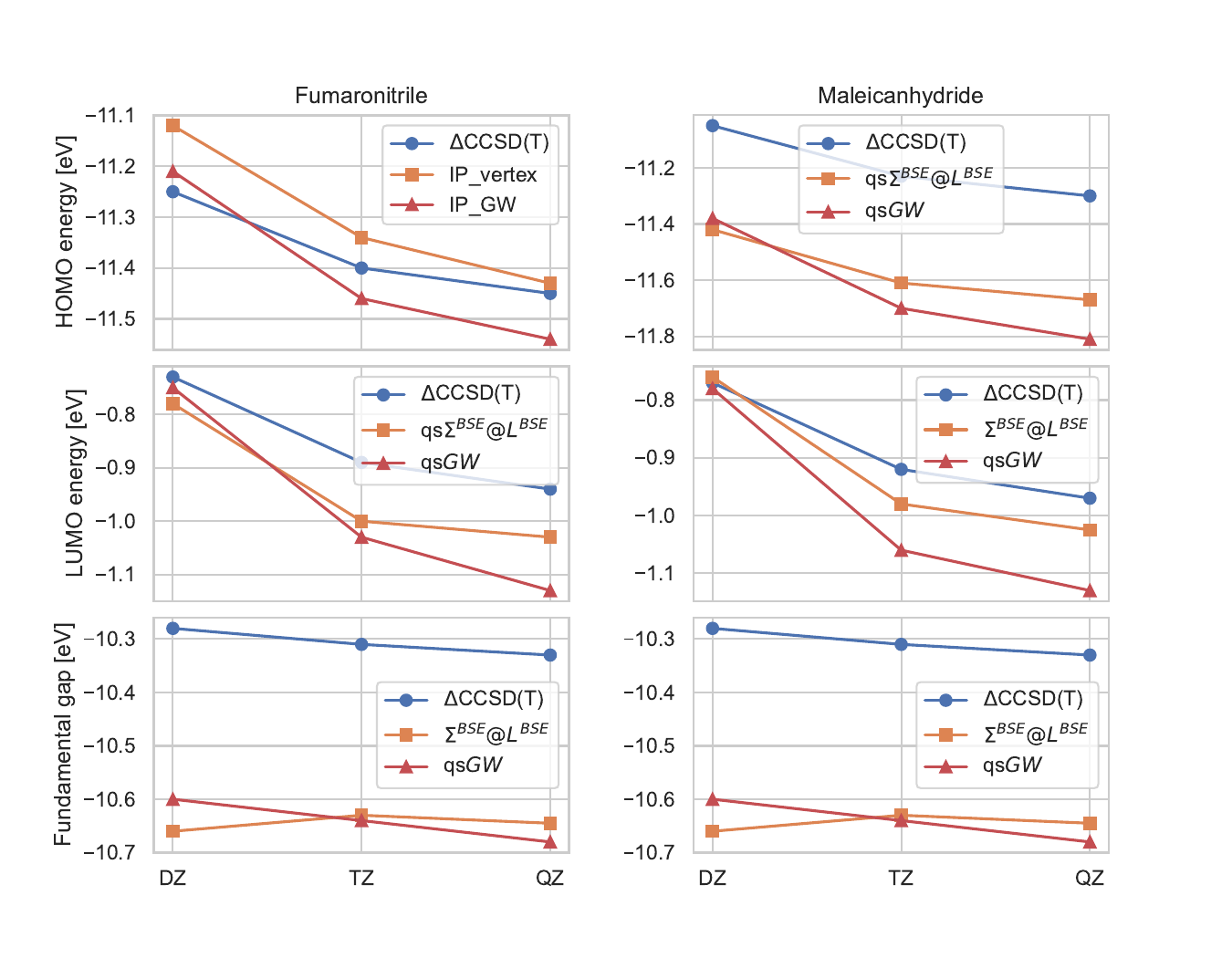}
    \caption{convergence of the HOMO energies (top), LUMO energies (middle), and fundamental gaps (bottom) of Fumaronitrile (left) and Maleic anhydride (right) with respect to the basis set (aug-cc-pVDZ to aug-cc-pVQZ) for $\Delta$CCSD(T)\cite{Richard2016}, qs$GW$ and qs$\Sigma^{BSE}@L^{BSE}$. All values are in eV}
    \label{fig:CBSconvergence_curves}
\end{figure}

In the following benchmarks, we prefer to use the pure $\Delta$CCSD(T) values as reference, and for a reliable comparison, we first investigate how $\Sigma^{BSE}@L^{BSE}$ converge to the CBS limit, and how this convergence compares to $\Delta$CCSD(T). HOMO energies, LUMO energies, and fundamental gaps of Maleic anhydride and Fumaronitrile calculated with all three methods using aug-cc-pVDZ to aug-cc-pVQZ are shown in Fig.~\ref{fig:CBSconvergence_curves}. 

The convergence of the HOMO energy is shown in the upper plot. $\Delta$CCSD(T) converges to the CBS limit faster than qs$\Sigma^{BSE}@L^{BSE}$ and qs$GW$ which show a similar rate of convergence. The LUMO energy of Fumaronitrile and the HOMO and LUMO energies of Maleic anhydride, converge much faster to the CBS limit with qs$\Sigma^{BSE}@L^{BSE}$ than with qs$GW$, almost as fast as with $\Delta$CCSD(T). Comparing the LUMO energies obtained with different basis sets is particularly insightful. For Fumaronitrile and Maleic anhydride, the qs$\Sigma^{BSE}@L^{BSE}$ and qs$GW$ LUMO energies agree with $\Delta$CCSD(T) within a few ten meV. However, this is partially an artifact of the non-converged basis set. At the QZ level, the difference in the qs$GW$ and $\Delta$CCSD(T) LUMO energies is about 200 meV for Fumaronitririle and 150 meV for Maleic anhydride. The agreement between qs$\Sigma^{BSE}@L^{BSE}$ and $\Delta$CCSD(T) remains better also for the larger basis sets. The difference in the LUMO energy of Fumaronitrile is 50 meV for DZ and 90 meV for QZ, and for Maleic anhydride the differences are 10 meV and 40 meV, respectively. We also notice that the QP gap between HOMO and LUMO is already converged sufficiently at the DZ level since the basis set errors for HOMO and LUMO are typically the same when augmented GTO basis sets are used.\cite{Blase2011}

\begin{figure}
    \centering
    \includegraphics[width=0.7\linewidth]{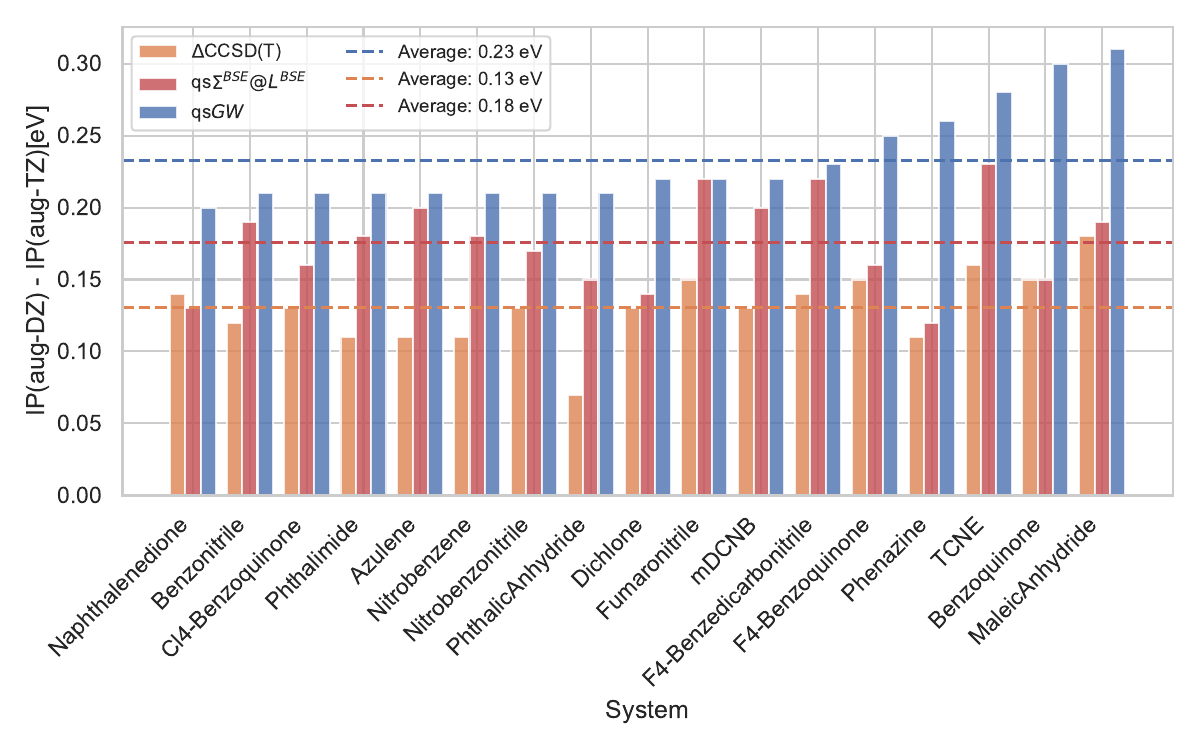}
    \caption{Difference in first IPs in eV calculated with the aug-cc-pVDZ and aug-cc-pVTZ basis sets for $\Delta$CCSD(T)\cite{Richard2016}, qs$GW$ and qs$\Sigma^{BSE}@L^{BSE}$.}
    \label{fig:CBSconvergence_bins}
\end{figure}

Fig.~\ref{fig:CBSconvergence_bins} shows the difference between the aug-cc-pVDZ and aug-cc-pVTZ first IPs for a subset of ACC24 for $\Delta$CCSD(T), qs$GW$ and qs$\Sigma^{BSE}@L^{BSE}$. For $\Delta$CCSD(T), the average difference in IPs obtained with both basis sets is 0.13 eV, while for qs$GW$, with 0.23 eV this difference is significantly larger. With 0.18 eV on average, qs$\Sigma^{BSE}@L^{BSE}$ converges to the CBS limit faster, not much slower than $\Delta$CCSD(T). The slow convergence of $GW$ to the CBS limit is due to the absence of exchange terms in the correlation part of the self-energy which converge to the CBS limit with opposite signs than the direct ones.\cite{Bruneval2024} This is an empirical observation, but has also been discussed in the context of CC theories.\cite{Irmler2019} Therefore, the vertex-corrections in qs$\Sigma^{BSE}@L^{BSE}$ lead to faster basis set convergence. 

The results of this subsection clearly show that care should be taken when $GW$ results are compared to high-level wave function-based methods in a small basis set. Comparing qs$GW$ to $\Delta$CCSD(T) in a DZ basis set will give a skewed picture of its accuracy. For our present study, we conclude that already at the DZ level qs$\Sigma^{BSE}@L^{BSE}$ can be faithfully compared to $\Delta$CCSD(T), and at the TZ level the basis set error should essentially play no role. The situation is different for qs$GW$, where a comparison at the DZ level will come with major basis set errors, and calculations of at least TZ level are necessary to assess its performance reliably. 

Therefore, all qs$GW$ QP energies shown in the following are obtained with the aug-cc-pVTZ basis set. The same holds for qs$\Sigma^{BSE}@L^{BSE}$, except for six molecules where we were not able to perform qs$\Sigma^{BSE}@L^{BSE}$/aug-cc-pVTZ calculations due to high memory demands. In these cases, we corrected the IPs and EAs with the average basis set errors of 0.18 eV shown in figure~\ref{fig:CBSconvergence_bins}. The error of this estimate should be smaller than 50 meV. The IPs and EAs of the 4 systems for which no aug-cc-pvTZ reference values are available are shifted by the $\Delta$MP2 values from Ref.~\citenum{Richard2016}. The same has been done for Dinitrobenzonitrile, where the aug-cc-pVTZ reference value seems to be incorrectly reported. This correction allows for a fair assessment of the method's accuracy for the complete ACC24 set. All individual QP energies are listed in the supporting information.

\subsubsection{Compensation of vertex corrections in $L$ and $\Sigma$}

\begin{figure}[hbt!]
    \centering
    \includegraphics[width=0.5\linewidth]{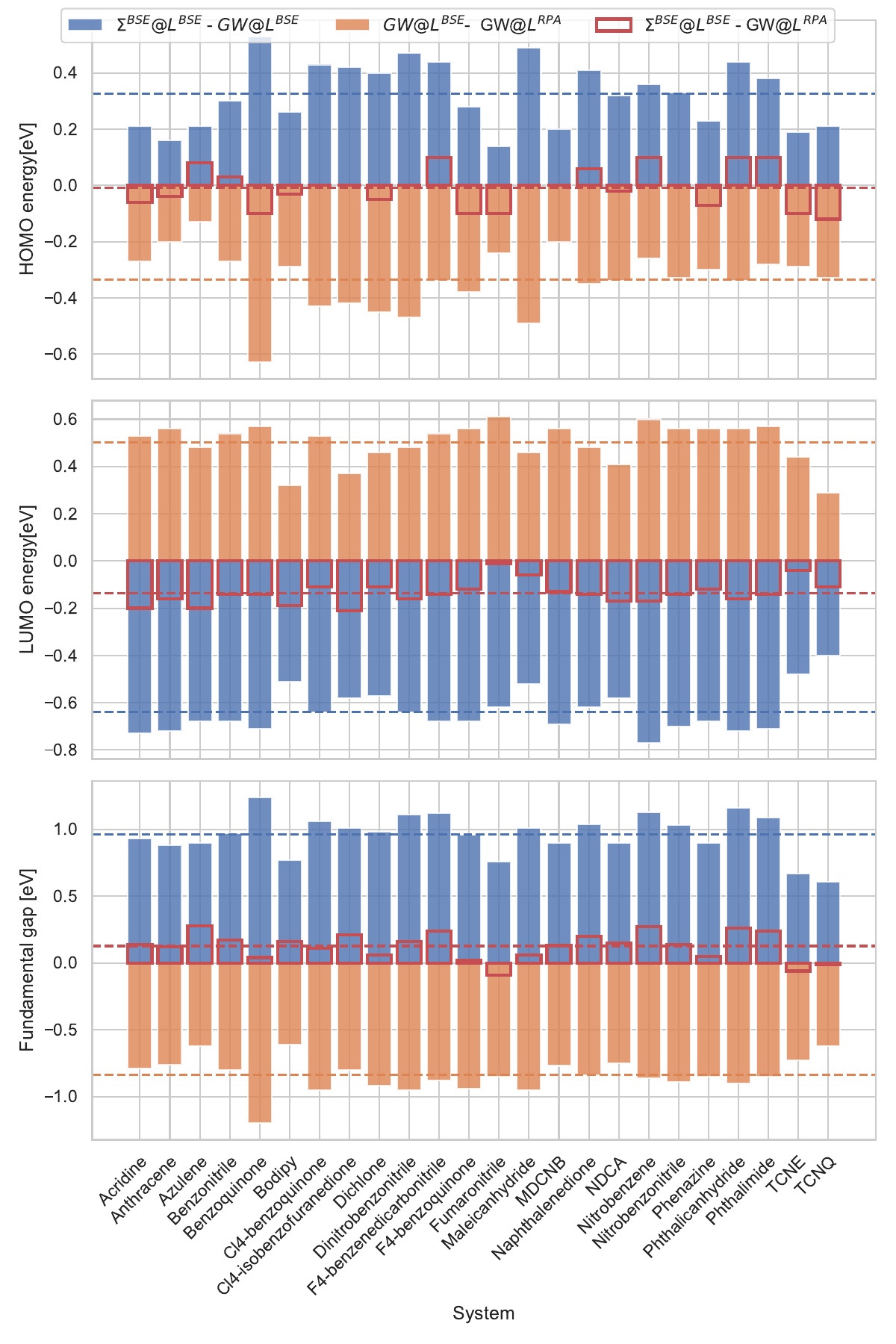}
    \caption{Vertex corrections in eV of the HOMO (top), LUMO (middle), and HOMO-LUMO gap (bottom) of the molecules in the ACC24.}
    \label{fig:compensation}
\end{figure}

Before comparing the (vertex-corrected) qs$GW$ calculations to $\Delta$CCSD(T), we show the effect of the individual vertex corrections in $L$ and $\Sigma$ in Fig.~\ref{fig:compensation}. The orange bars show the magnitude of the vertex correction beyond RPA in $L$ (corresponding to the green dots in Fig.~\ref{fig:benchmark}) and the blue bars show the magnitude of the vertex correction in $\Sigma$ beyond $GW$ (corresponding to the blue dots in Fig.~\ref{fig:benchmark}). The red boxes are the blue and orange bars' sums, showing the difference between $GW$@RPA and $\Sigma^{BSE}@L^{BSE}$.  As observed previously,\cite{Forster2024} the vertex corrections in $\Sigma$ and $L$ largely compensate for the HOMO. However, they substantially increase the LUMO (for which the magnitudes of the vertices in $L$ and $\Sigma$ are also much larger), widening the qs$GW$ HOMO-LUMO gap. This has also been observed by \citet{Gruneis2014} This is likely an artifact of the static vertex. Accounting for the dynamics of the vertex (at least within sc$GW$) should close the fundamental gap.\cite{Kutepov2016} 

\subsubsection{Accuracy of vertex-corrected qs$GW$}

\begin{figure}
    \centering
    \includegraphics[width=\linewidth]{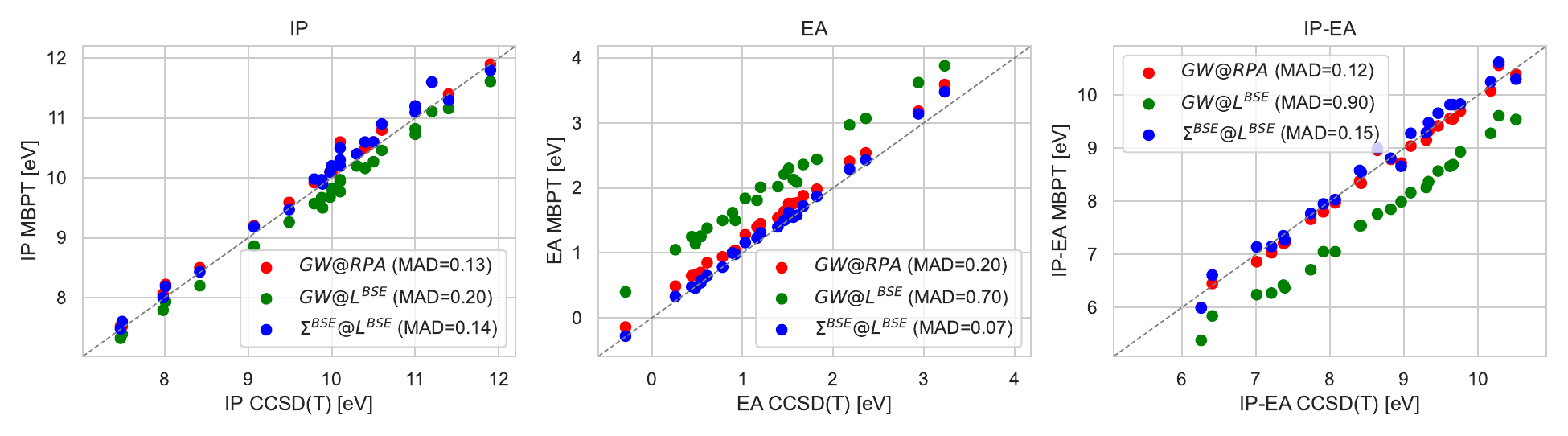}
    \caption{Deviations of the molecules in the ACC24 set to $\Delta$CCSD(T) for quasi-particle self-consistent $GW$@RPA, $GW$@$L^{BSE}$ and $\Sigma^{BSE}$@$L^{BSE}$ for IPs (left), EAs (middle) and fundamental gaps (right). All values are in eV.}
    \label{fig:benchmark}
\end{figure}

After discussing the basis set errors of the different approximations, we are in a position to compare qs$GW$@RPA, qs$GW$@$L^{BSE}$ and qs$\Sigma^{BSE}$@$L^{BSE}$ against the $\Delta$CCSD(T) reference values from \citet{Richard2016} for the ACC24 set. Fig.~\ref{fig:benchmark} plots the IPs, EAs, and fundamental gaps obtained with all three MBPT approximations against $\Delta$CCSD(T). In the supporting information, we show the same plot, but with all values calculated with the aug-cc-pVDZ basis set. Comparing both plots shows, that a benchmark using aug-cc-pVDZ would lead to a skewed picture of the respective methods' accuracy. 

We first discuss the IPs. As can be seen from the upper panel in Fig.~\ref{fig:compensation}, the vertex corrections in $L$ and $\Sigma$ in qs$\Sigma^{BSE}$@$L^{BSE}$ cancel almost completely. Therefore, the differences between the qs$\Sigma^{BSE}$@$L^{BSE}$ and qs$GW$@RPA IPs are small. Consequently, with MADs of 0.13 and 0.14 eV, respectively, the accuracy of both methods in predicting IPs is very similar. For EAs, the vertex in $L$ dominates the vertex in $\Sigma$ and therefore the qs$\Sigma^{BSE}$@$L^{BSE}$ EAs are much lower than the qs$GW$@RPA ones. As shown in the middle plot in Fig.~\ref{fig:benchmark}, this massively improves the agreement with the $\Delta$CCSD(T) reference values. qs$\Sigma^{BSE}$@$L^{BSE}$ gives an excellent MAD of 0.07 eV, while with 0.20 eV, qs$GW$@RPA performs significantly worse. Consequently, qs$\Sigma^{BSE}$@$L^{BSE}$ gives much larger fundamental gaps than qs$GW$@RPA. qs$GW$@RPA underestimates the $\Delta$CCSD(T) gaps, but qs$\Sigma^{BSE}$@$L^{BSE}$ tends to overcorrect them. In the end, with 0.12 and 0.15 eV, respectively, both methods offer very similar accuracy. Finally, we also comment on the performance of qs$GW$@$L^{BSE}$. The IPs are of surprisingly good quality. However, as already anticipated, due to the missing vertex correction in $\Sigma$, qs$GW$@$L^{BSE}$ overestimates EAs, and therefore massively underestimates band gaps.

In Ref.~\citenum{Forster2022} we have already performed qs$GW$@RPA calculations for the ACC24 set and obtained MADs for IPs, EAs, and fundamental gaps as 0.09 eV, 0.14 eV, and 0.13 eV. In Ref.~\citenum{Forster2022} we have obtained all results by extrapolating the QP energies to the CBS limit using STO-type basis sets of TZ and QZ quality and used the $\Delta$CCSD(T)/CBS values of Ref.~\citenum{Richard2016}. Our old results are more accurate since they are basis set limit extrapolated, but the results presented here are more suitable to compare the qs$GW$ charged excitation energies against the reference values of Ref.~\citenum{Richard2016}. Therefore, we must dampen our positive conclusions on the high accuracy of qs$GW$@RPA a bit. Here, we find the performance of qs$GW$@RPA to be slightly worse for IPs and EAs, but its performance for fundamental gaps is still excellent. 

\begin{table}[hbt!]
    \centering
    \begin{tabular}{l
    S[table-format=-3.2]
S[table-format=-3.2]
S[table-format=-3.2]l}
\toprule
Method & {MAD [IP]} & {MAD [EA]} & {MAD [gap]} & Reference \\
\midrule
qs$GW$@RPA                        & 0.13 & 0.20 & 0.12 & This work \\
qs$\Sigma^{BSE}$@$L^{BSE}$        & 0.14 & 0.07 & 0.15 & This work \\
$G_0W_0$@LRC-$\omega$PBE          & 0.13 & 0.18 &      & Ref.~\citenum{Knight2016}\\ 
$G_0W_0$@LC-$\omega$PBE (tuned)   & 0.09 & 0.13 & 0.13 & Ref.~\citenum{Gallandi2016}\\
$G_0W_0$@OTRSH                    & 0.09 & 0.07 & 0.14 & Ref.~\citenum{Bruneval2024}\\
ADC(3)                            & 0.12 & 0.16 &      & Ref.~\citenum{Dolgounitcheva2016}\\
$\omega$LH22t                     & 0.15 & 0.18 & 0.23 & Ref.~\citenum{Furst2023} \\
\bottomrule
    \end{tabular}
    \caption{MADs in eV for the ACC24 set for IPs, EAs, and fundamental gaps obtained with different accurate methods.}
    \label{tab:MADs}
\end{table}

Finally, we compare the performance of qs$GW$@RPA and qs$\Sigma^{BSE}$@$L^{BSE}$ against other $GW$ approaches. MADs obtained with different methods are collected in table~\ref{tab:MADs}. \citet{Knight2016} have benchmarked the performance of several $GW$ methods for the ACC24 set and found $G_0W_0$@LRC-$\omega$PBE to perform best, with MADs of 0.13 eV for IPs, and 0.18 eV for EAs. This is very much comparable to our current qs$GW$@RPA results which shows that qs$GW$ can compete in accuracy with the best one-shot $G_0W_0$ methods. In Ref.~\citenum{Bruneval2024}, MADs of 0.09 eV for the IPs, 0.07 eV for the EAs and 0.14 eV for the fundamental gaps have been reported using $G_0W_0$ with the optimally-tuned range-separated hybrid (OTRSH) strategy of Ref.~\citenum{McKeon2022} ($G_0W_0$@OTRSH). Similar accuracy is achieved for differently tuned $G_0W_0$@OTRSH functionals.\cite{Gallandi2016} $G_0W_0$@OTRSH can also be understood as a self-consistent $GW$ approach since multiple $G_0W_0$ calculations have to be performed for different range-separation parameters until the $G_0W_0$ correction vanishes. It is worth pointing out that the tuning procedure applied in these works might become problematic for molecules much larger than the ones studied here.\cite{Gallandi2016} 
qs$\Sigma^{BSE}$@$L^{BSE}$ performs equally well for EAs and fundamental gaps. Even though its performance for IPs is worse, it still outperforms most $GW$-based methods. qs$\Sigma^{BSE}$@$L^{BSE}$ also outperforms third-order algebraic diagrammatic construction for EAs [ADC(3)]\cite{VonNiessen1984}, for which MADs of 0.12 eV for IPs and 0.16 eV for EAs have been reported for the same set.\cite{Dolgounitcheva2016} It should be noted that small differences in the MADs reported in table~\ref{tab:MADs} should not be over-interpreted since basis set incompleteness errors can not be excluded. For instance, the CBS limit for the ACD(3) results has been obtained by extrapolating from aug-cc-pvDZ and aug-cc-pvTZ basis sets\cite{Dolgounitcheva2016} which is generally error-prone.\cite{Jensen2013}

\subsection{Neutral excitations}

We next calculate the neutral CT excitation energies of the larger molecules in the QUEST \#6 database.\cite{Loos2021} 
We notice, that BSE excitation energies of the QUEST \# 3 database have recently been calculated by \citet{Waide2024} using the $\Sigma^{TDHF}$@HF and $\Sigma^{BSE}$ (called $\Sigma^{scTDHF}$@HF in their work) starting points. 
We focus here on QUEST \#6 since the systems in this database are similar to the ones in ACC24. For instance, Nitrobenzene, Benzonitrile, and Azulene are part of both databases. We therefore can assess whether the accuracy of qs$\Sigma^{BSE}@L^{BSE}$ for fundamental gaps carries over to neutral excitations.

\begin{figure}[hbt!]
    \centering
    \includegraphics[width=0.5\linewidth]{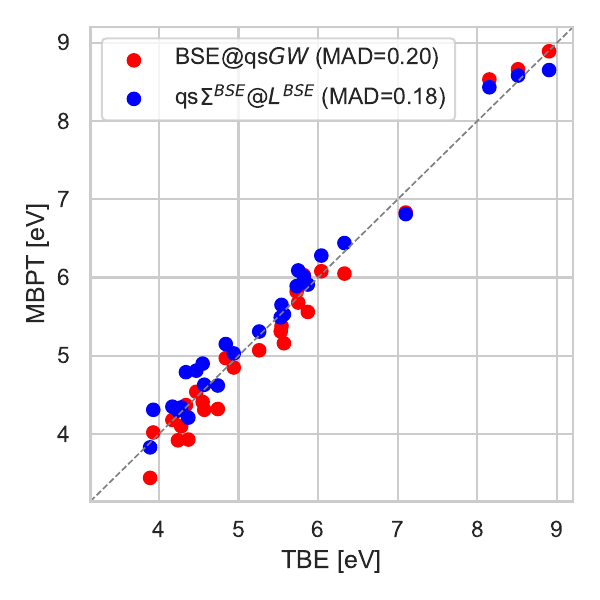}
    \caption{Comparison of neutral charge-transfer excitations in eV for the QUEST \# 6 database calculated with BSE@qs$GW$ and qs$\Sigma^{BSE}@L^{BSE}$ against theoretical best estimates (TBE).\cite{Loos2021}}
    \label{fig:quest6}
\end{figure}

In Fig.~\ref{fig:quest6}, we compare the CT excitation energies obtained with BSE@qs$GW$ and qs$\Sigma^{BSE}@L^{BSE}$ against the theoretical best estimates (TBE) of Loos et al.\cite{Loos2021} As expected, the excitation energies strongly correlate with the fundamental QP gap. The wider gap of qs$\Sigma^{BSE}@L^{BSE}$ with respect to qs$GW$ also results in higher CT excitation energies. Only for the $A^{\prime \prime}$ excitations in $\beta$-dipeptide and dipeptide, qs$\Sigma^{BSE}@L^{BSE}$ gives lower excitation energies than qs$GW$. In both cases, the qs$\Sigma^{BSE}@L^{BSE}$ fundamental gap is also lower than the qs$GW$ one.  

For the QUEST~\#6 database, the BSE@qs$GW$ and qs$\Sigma^{BSE}@L^{BSE}$ neutral excitations are of the same quality as the ones calculated by \citet{Loos2021} using BSE@ev$GW$@PBE0. \citet{Loos2021} have also benchmarked the accuracy of TD-DFT with different range-separated hybrid functionals and found them to perform slightly better than BSE@$GW$ on average for this particular class of systems. One should notice, that this good performance was only attained by two functionals (CAM-B3LYP\cite{Yanai2004} and $\omega$B97X-D\cite{Chai2008a}), while two other functionals ($\omega$B97X\cite{Chai2008} and LRC-$\omega$HPBE\cite{Rohrdanz2009}) perform much worse.\cite{Loos2021} A similar observation was for instance made in Ref.~\citenum{Forster2024}. Since it is not clear a priori which functional will perform best for what system, this is a clear disadvantage of TD-DFT@RSH compared to BSE@$GW$. We however also mention that there exist approaches to non-empirically tune the range-separation parameter in RSHs\cite{Stein2010, Refaely-Abramson2011, Refaely-Abramson2012, Kronik2012} which alleviate this issue a bit.

The observation that qs$\Sigma^{BSE}@L^{BSE}$ does not outperform BSE@qs$GW$ is not unexpected. The former method does not introduce any new diagrams in the BSE. qs$\Sigma^{BSE}@L^{BSE}$ does only provide different screening and different QP energies, but the QP differences, the quantity entering excited state calculations, are of similar quality. To improve over BSE@$GW$, one likely has to go beyond the first-order vertex in the BSE, as has already been attempted by some authors.\cite{Yamada2022, Yamada2023, Monino2023}  

\section{\label{sec:Conclusions}Conclusions}
We have implemented and tested the $\Sigma^{BSE}@L^{BSE}$ self-energy in a quasi-particle self-consistent framework.\cite{Faleev2004, VanSchilfgaarde2006, Kotani2007, bruneval_springer2014} $\Sigma^{BSE}@L^{BSE}$ goes beyond $GW$ by adding statically screened particle-hole ladders to the response function $L$ and the self-energy, going beyond (diagrammatic) approaches that solve the BSE for $L$ but do not add vertex corrections to $\Sigma$ directly.\cite{Cunningham2018, Cunningham2024} Adding vertex corrections to $L$ only leads to a disastrous performance for molecular EAs and even worse fundamental gaps. Vertex corrections in $\Sigma$ are needed to cancel the strong electron-hole attraction induced through the vertex in $L$. While vertex corrections in $L$ and $\Sigma$ largely cancel, they increase LUMO energies significantly, and therefore open the fundamental gap. While this leads to a slight overestimation of the gap compared to $\Delta$CCSD(T), the performance of qs$\Sigma^{BSE}@L^{BSE}$ is excellent. It retains the great accuracy of qs$GW$ for fundamental gaps and IPs, but greatly improves the EAs. These observations agree with Ref.~\citenum{Vlcek2019}, where the non-screened version of the present self-energy approximation was benchmarked with a HF starting point. An advantage of qs$\Sigma^{BSE}@L^{BSE}$ over qs$GW$ is its faster convergence to the CBS limit which will be important in practice. We have also assessed qs$\Sigma^{BSE}@L^{BSE}$ for neutral CT excitations, and found it to perform similarly to qs$GW$.

The $\Sigma^{BSE}@L^{BSE}$ self-energy has only been introduced very recently.\cite{Patterson2024, Patterson2024a,Forster2024} Its performance is similar to the self-energy obtained by replacing the statically screened terms with unscreened ones.\cite{Maggio2017, Maggio2018, Vlcek2019, Vacondio2024, Forster2024} In both approximations, the very same vertex is consistently added to $L$ and $\Sigma$, which is important for systematic improvements beyond $GW$.\cite{Bruneval2024, Forster2024} Using the screened interaction instead, as commonly done in BSE@$GW$ calculations,\cite{Blase2018} becomes important for larger molecules.\cite{Forster2024} These recently introduced self-energy approximations are an important step toward robust and diagrammatically motivated approximations to the self-energy beyond the GWA.

At this stage, we see numerous promising avenues for future research that we hope will be pursued soon. Going beyond the quasi-particle approximation and introducing the $\Sigma^{BSE}@L^{BSE}$ self-energy in molecular sc$GW$ calculations would be a logical extension of the current work. Due to the relatively strong QP renormalization in solids, it is often claimed that sc$GW$ calculations should always be combined with vertex corrections.\cite{bruneval_springer2014, Rohlfing2023} Using sc$GW$ calculations in small atoms and molecules where screening effects are weaker seems to be well justified from a theoretical perspective.\cite{Stan2006, Stan2009, bruneval_springer2014} While sc$GW$ calculations are rarely performed and few implementations for molecules exist,\cite{Rostgaard2010, Koval2014, Caruso2012, Caruso2013, Abraham2024, Iskakov2024} very recently Zgid and coworkers reported good agreement with experimental data for the GW27 and the SOC81 sets of molecular IPs.\cite{Scherpelz2016, Harsha2024, Abraham2024} It would be worthwhile to investigate whether sc$GW$ calculations can be systematically improved with vertex corrections, as has been done for solids.\cite{Kutepov2016, Kutepov2017} As a promising step in this direction, first-order vertex corrections to the $GW$ self-energy have recently been implemented by Zgid and co-workers in an sc$GW$ framework and applied to the calculation of exchange couplings in solids and molecules.\cite{Pokhilko2024a}

Including dynamical vertex effects would be another possible extension of this work. Kutepov could show that including the dynamical vertex in $L$ and $\Sigma$ closes the sc$GW$ band gaps, while the static vertex opens them further.\cite{Kutepov2016} The challenge with dynamical vertices lies in the fact that the BSE \cref{response} does not admit a closed-form expression as the $L$ on the left exhibits a different frequency dependence from the $L$ on the right. Consequently, the equation can no longer be solved through diagonalization.\cite{Romaniello2009b, Sangalli2011} \citet*{Kuwahara2016} have combined the dynamical first-order vertex correction to the RPA polarizability with the dynamical $G3W2$ self-energy. While this scheme uses consistent vertices, the first-order vertex corrections in $L$ and $\Sigma$ cancel almost completely\cite{Bobbert1994, DeGroot1996, Forster2024} and therefore only little improvement over qs$GW$ can be expected. Kutepov went beyond first-order in the dynamical vertex and used an iterative procedure to sum the dynamical BSE \cref{response} order-by-order.\cite{Kutepov2016} One can also think of schemes where the first-order dynamical vertex corrections of \citet{Kuwahara2016} are combined with the infinite-order static correction both for $\Sigma$ and $L$.

Vertex corrections beyond the first order. would be another extension of this work. \citet{Monino2023} and Yamada \textit{et. al.}\cite{Yamada2022, Yamada2023} have found second-order vertex diagrams to change BSE@$GW$ neutral excitation energies significantly. These studies considered only the two of the six second-order vertex diagrams which arise from the variation of $W$ with respect to $G$. It remains an open question whether further cancellations occur among all six diagrams. It is also not known if the second-order vertex diagrams in $\Sigma$ and $L$ compensate in the same way as the 1st-order diagrams beyond $GW$.\cite{Bobbert1994, DeGroot1996, Ummels1998, Forster2024} Higher-order vertex corrections Have been investigated by \citet{Mejuto-Zaera2022a} for the Hubbard dimer, and there also exists a recent proposal along these lines by Cunningham.\cite{Cunningham2024} Implementing and testing such higher-order corrections for molecules would be important to improve the quality of neutral excitation energies. Second-order diagrams in the kernel beyond the ones arising from the variation of $W$ with respect to $G$ could potentially improve triplet excitations, which are only badly described with BSE@$GW$.\cite{Rangel2017}

\begin{acknowledgement}
The author acknowledges supercomputer facilities at SURFsara sponsored by NWO Physical Sciences, with financial support from The Netherlands Organization for Scientific Research (NWO) and fruitful discussions with Fabien Bruneval.
\end{acknowledgement}

\begin{suppinfo}
All QP energies and neutral excitation energies calculated in this work,
\end{suppinfo}

\begin{tocentry}
\includegraphics[width=\textwidth]{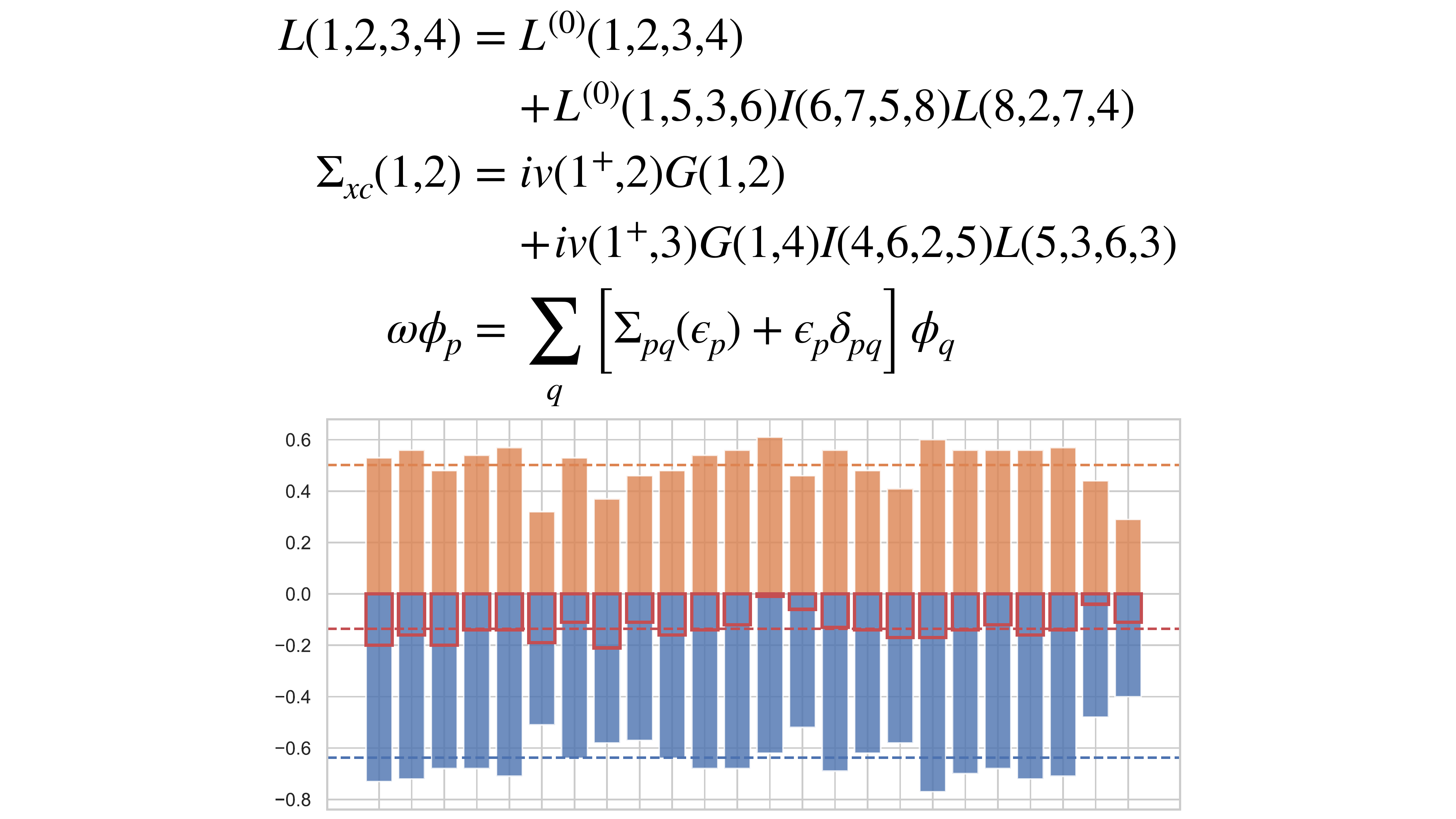}
\end{tocentry}

\appendix

\bibliography{all.bib, Fabien.bib}

\begin{mcitethebibliography}{152}
\bibitem[Not()]{Note-1}
Similar to other electronic structure methods, for instance, Hartree--Fock (HF) theory, there can of course be multiple solutions corresponding to different local minima which may be reached from different starting points\relax
\mciteBstWouldAddEndPuncttrue
\mciteSetBstMidEndSepPunct{\mcitedefaultmidpunct}
{\mcitedefaultendpunct}{\mcitedefaultseppunct}\relax
\EndOfBibitem
\end{mcitethebibliography}

\end{document}